\documentclass[sn-nature,Numbered]{sn-jnl}%
\usepackage{graphicx}%
\usepackage{multirow}%
\usepackage{amsmath,amssymb,amsfonts}%
\usepackage{amsthm}%
\usepackage{mathrsfs}%
\usepackage[title]{appendix}%
\usepackage{xcolor}%
\usepackage{textcomp}%
\usepackage{manyfoot}%
\usepackage{booktabs}%
\usepackage{algorithm}%
\usepackage{algorithmicx}%
\usepackage{algpseudocode}%
\usepackage{listings}%
\usepackage{hyperref}
\usepackage{subfig}
\hypersetup{hidelinks,
	colorlinks=true,
	allcolors=black,
	pdfstartview=Fit,
	breaklinks=true}

\begin{document}
\title{Explanations of Classifiers Enhance Medical Image Segmentation via End-to-end Pre-training}
%
%
%
%
%
%
\author[1]{\fnm{Jiamin} \sur{Chen}}\email{chenjiamin01@baidu.com}
\author[1]{\fnm{Xuhong} \sur{Li}}\email{jacqueslixuhong@gmail.com}
\author[2]{\fnm{Yanwu} \sur{Xu}}\email{ywxu@ieee.org}
\author[3]{\fnm{Mengnan} \sur{Du}}\email{mengnan.du@njit.edu}
\author[1]{\fnm{Haoyi} \sur{Xiong}}\email{haoyi.xiong.fr@ieee.org}
\affil[1]{\orgdiv{Big Data Lab}, \orgname{Baidu Inc}, \orgaddress{\street{Haidian}, \postcode{100085}, \state{Beijing}, \country{China}}}
\affil[2]{\orgdiv{HDMI Lab}, \orgname{South China University of Technology}, \orgaddress{\city{Guangzhou}, \state{Guangdong}, \country{China}}}
\affil[3]{\orgdiv{Department of Data Science}, \orgname{New Jersey Institute of Technology}, \orgaddress{\street{Dr Martin Luther King Jr Blvd}, \state{Newark}, \postcode{07102}, \country{NJ}}}
\abstract{
Medical image segmentation aims to identify and locate abnormal structures in medical images, such as chest radiographs, using deep neural networks. These networks require a large number of annotated images with fine-grained masks for the regions of interest, making pre-training strategies based on classification datasets essential for sample efficiency. Based on a large-scale medical image classification dataset, our work collects explanations from well-trained classifiers to generate pseudo labels of segmentation tasks. Specifically, we offer a case study on chest radiographs and train image classifiers on the CheXpert dataset to identify 14 pathological observations in radiology. We then use Integrated Gradients (IG) method to distill and boost the explanations obtained from the classifiers, generating massive diagnosis-oriented localization labels (DoLL). These DoLL-annotated images are used for pre-training the model before fine-tuning it for downstream segmentation tasks, including COVID-19 infectious areas, lungs, heart, and clavicles. Our method outperforms other baselines, showcasing significant advantages in model performance and training efficiency across various segmentation settings.

}

\maketitle              

\section{Introduction}
Medical image segmentation, crucial for identifying and localizing abnormal structures in medical images, such as diagnosing pneumonia, heart failure and hiatal hernia from chest radiographs (a.k.a chest X-ray or CXR), has significantly benefited from deep neural networks (DNNs)~\cite{rajaraman2021improved,DBLP:journals/cbm/TahirCKRQKKIRAM21}. Many studies have been proposed towards the medical image segmentation for chest X-rays through supervised deep learning~\cite{DBLP:conf/miccai/RonnebergerFB15,DBLP:journals/pami/BadrinarayananK17,liu2022automatic,rajaraman2021improved}. However, all these work rely on a large number of annotated images with fine-grained masks for the regions of interest, posing a challenge regarding sample efficiency. Consequently, pre-training strategies leveraging image classification datasets have become essential to address this issue~\cite{DBLP:conf/eccv/XieR18,DBLP:conf/miccai/LiaoXWMLLCHD22}.



To effectively pre-train a segmentation model for medical image segmentation, one can either utilize natural image classification datasets, such as ImageNet~\cite{DBLP:conf/cvpr/DengDSLL009} and Grayscale ImageNet~\cite{DBLP:conf/eccv/XieR18}, or leverage datasets specifically curated for medical image classification~\cite{DBLP:conf/miccai/LiaoXWMLLCHD22,DBLP:conf/eccv/KalaposG22,DBLP:conf/eccv/HuangLBK18,DBLP:journals/titb/LiangGZXWKJ22}. The backbone of classifiers trained on such datasets can then be adapted to serve as pre-trained weights for the segmentation model (backbone+segmentation module), facilitating improved performance in medical image segmentation. 
However, there still remain some challenges for the above studies. 
For the backbone pre-training, as they only initialize the backbone weights and leave the segmentation module randomly initialized, it still needs much annotated data for fine-tuning. Some works~\cite{DBLP:conf/cvpr/ZhaoSQWJ17,li2023distilling} propose the end-to-end pre-training strategies with large-scale annotated datasets of natural images, such as Microsoft COCO~\cite{DBLP:conf/eccv/LinMBHPRDZ14}. However, for CXR images, there does not exist so far any large annotated dataset. Unlike natural images with more morphological details to annotate, the semantic information inside chest X-rays is non-obvious and can be perceived under different views~\cite{DBLP:journals/jvcir/WenCDZ21}.

\begin{figure}[htbp]
    \centering
    \subfloat[Generating Diagnosis-oriented Localization Labels (DoLL)]{\includegraphics[width=0.85\textwidth]{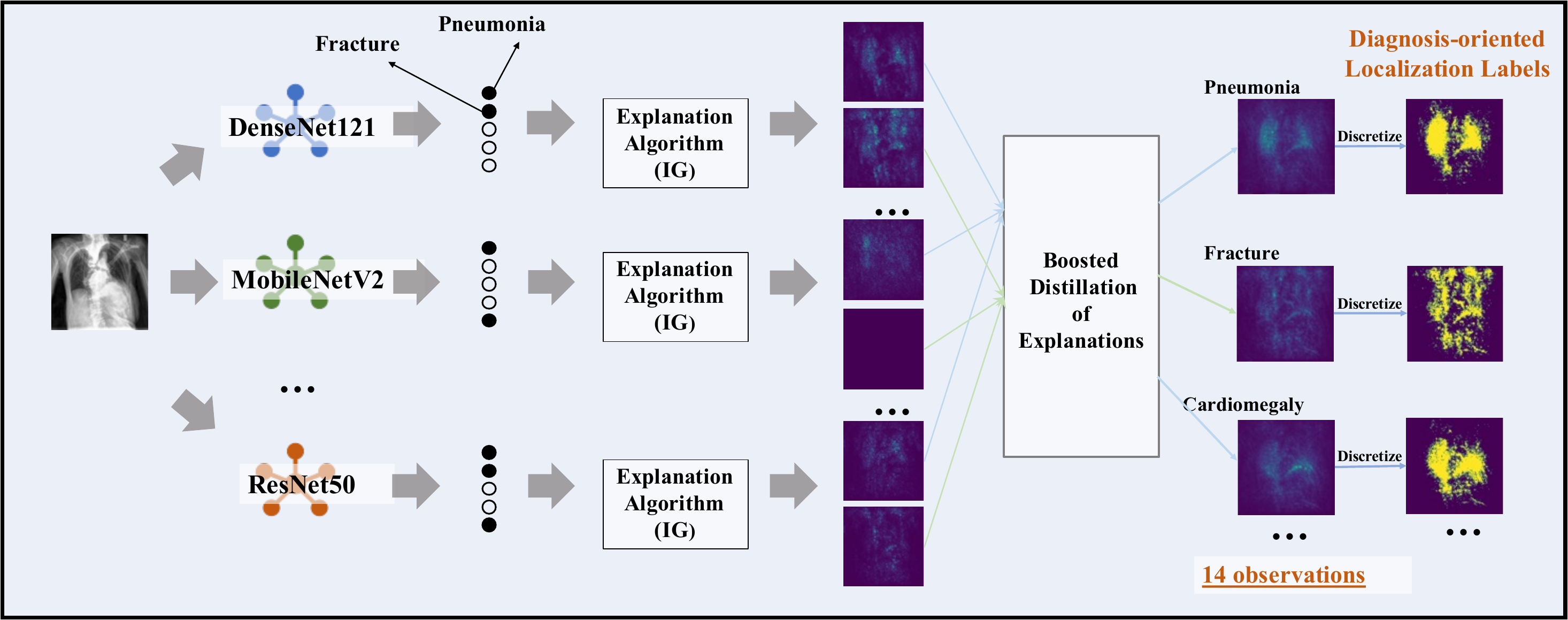}    \label{pipeline}}\\
    \subfloat[End-to-end Pre-training and Fine-tuning processes]{\includegraphics[width=0.6\textwidth]{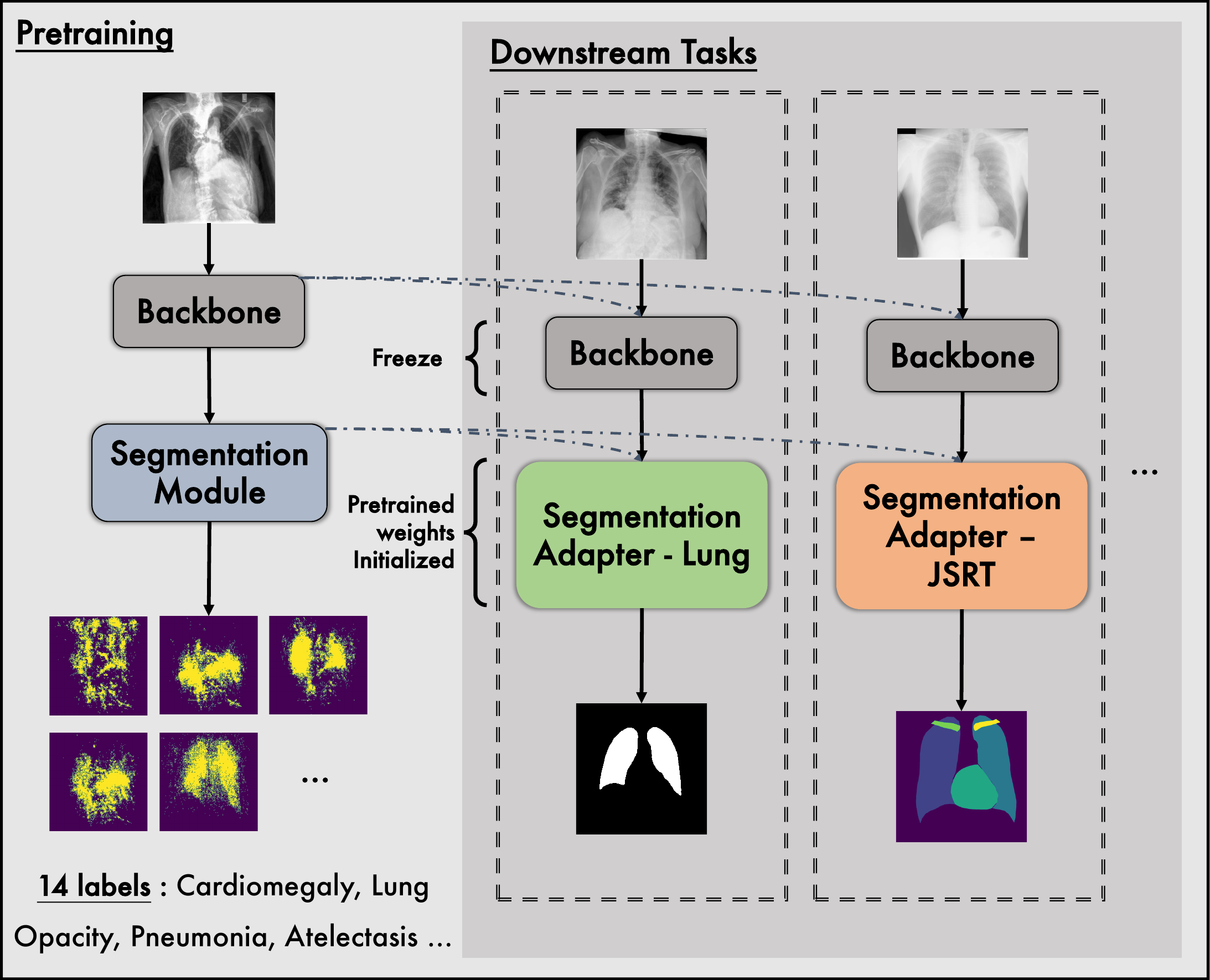}    \label{training}}
    \caption{An Illustration of DoLL-based Pre-training Approach based on CXR Classification Datasets, Classifiers, and Explainers}\label{overall}
\end{figure}

On the other hand, recent studies in explainable artificial intelligence (XAI) have demonstrated the feasibility of using activation maps, saliency maps, or even input gradients of a model to interpret the decisions made by DNNs~\cite{li2022interpretable}. Clinicians and researchers adopted these interpretations to explain the diagnostic results to patients or even advance the precise diagnostics that can be undetectable to the naked eye~\cite{essemlali2020understanding,el2021multilayer,borys2023explainable}. More specifically, our previous studies found the explanation result of an image classifier would be spatially closed to the location of visual objects for classification in the image~\cite{li2023cross}. Through aggregating the explanation results from multiple well-trained DNNs on the same image, it is possible to obtain a cross-model consensus of explanations via pixel-wise majority voting~\cite{li2023cross}. Such cross-model consensus of explanations could be further used as the \emph{Pseudo Semantic Segmentation Label} of the image (which was originally annotated for classification) to improve the performance of segmentation tasks through pre-training~\cite{li2023distilling,liu2023pixmim}.

In this study, we introduce a novel strategy, the Diagnosis-oriented Localization Labels (DoLL), to seamlessly pre-train deep neural networks (DNNs) for medical image segmentation tasks using only classification datasets. This approach is exemplified through a suite of chest X-ray (CXR) segmentation tasks. As illustrated in our framework shown in Fig.~\ref{overall}, we refine and enhance the explanatory power of image classifiers across 14 clinically relevant chest radiographic observations, adhering to the Fleischner Society's glossary standards~\cite{hansell2008fleischner}. Distinct from the traditional segmentation pre-training-fine-tuning workflow, which typically initializes only the model's backbone, our method endows the entire model, including the segmentation module, with superior pre-trained weights. This setup significantly streamlines the fine-tuning process by concentrating updates solely within the segmentation module, referred to as downstream segmentation adapters. Our contributions are manifold:
\begin{enumerate}
    \item We devise the DoLL method for automatically annotating chest X-rays, by distilling and boosting the explanations of classifiers trained on the frontal-view CXRs on CheXpert~\cite{irvin2019chexpert} for 14 pathological observations. With this method, we elaborate a large-scale annotated dataset \textbf{CheXpert-DoLL} for pre-training. It supports many downstream CXR segmentation tasks concerning bones, heart, lung, pleura, and the abnormalities within these regions. The \textbf{CheXpert-DoLL} dataset is publicly available at \url{http://somewhere}.
    \item We present an end-to-end pre-training method for chest X-ray segmentation based on \textbf{CheXpert-DoLL}. By associating X-ray pixels with meaningful categories according to clinical practice, this end-to-end pre-training method leaves the whole segmentation model, including both backbone and segmentation module, with a deep and complete understanding to chest X-ray images. In contrast to conventional approaches that apply pre-training solely to the backbone weights, our method allows for the use of pre-trained weights for both the backbone and the segmentation module. This comprehensive pre-training process significantly enhances the adaptation of the model to downstream segmentation tasks during fine-tuning. 
    
    \item We conduct extensive experiments with different network architectures on various downstream settings including lung segmentation, COVID-19 infection segmentation, and multi-organ segmentation of lungs, heart and clavicles. Both the experimental results and training efficiency demonstrate the great advantages of pre-training with DoLL against the other self-supervised pre-training methods, such as MoCo-CXR~\cite{DBLP:conf/midl/SowrirajanYNR21} and MoCo-v2~\cite{DBLP:journals/corr/ChenPSA17}, and pretrained backbones on ImageNet, Grayscale ImageNet~\cite{DBLP:conf/eccv/XieR18} and CheXpert~\cite{irvin2019chexpert}.
    The CheXpert-DoLL dataset and the pretrained segmentation models will be released publicly for future usage.
\end{enumerate}

\section{Related Works}
Pre-training strategies and the subsequent fine-tuning have become a crucial area of focus in recent research~\cite{DBLP:conf/cvpr/SinghGARGK0GDM22}, with implications extending across various domains including medical imaging~\cite{xie2018pre,wen2021rethinking,liao2022muscle}. While fully supervised pre-training and self-supervised pre-training have seen significant advancements~\cite{xie2018pre,wen2021rethinking,wolf2023self}, the territory of weakly-supervised pre-training remains comparatively underexplored. 

In the realm of visual recognition systems, Singh et al. revisit the potential of weakly-supervised pre-training, utilizing models that are pre-trained with hashtag-based supervision and demonstrating that such approaches can rival the performance of self-supervised methods~\cite{DBLP:conf/cvpr/SinghGARGK0GDM22}. Furthermore, Ghadiyaram et al. explored the application of large-scale weakly supervised pre-training for video models within the context of action recognition, showcasing the versatility of weak supervision beyond static imagery~\cite{DBLP:conf/cvpr/GhadiyaramTM19}.
Within medical imaging, despite the proposition of several weakly-supervised techniques in recent years, challenges persist. Some methods prove to be incompatible with segmentation tasks~\cite{zhou2020comparing}, while others may impose prohibitive time and resource demands for effective large-scale pre-training~\cite{viniavskyi2020weakly}. Liao~\emph{et al.} proposed a weakly-supervised pre-training strategy that combines unsupervised contrastive learning and supervised continual learning tasks into one pre-training pipeline for advanced performance based on X-ray images~\cite{liao2022muscle}.

Our work follows the concept of ``Learning from explanations'', which offers a novel perspective—incorporating insights gleaned from Explainable AI (XAI) techniques into the learning process to enhance the model's interpretability and accuracy, particularly for segmentation tasks. Class Activation Maps (CAM) paved the way by employing explanatory visual cues to locate discriminative regions within classification models~\cite{DBLP:conf/cvpr/ZhouKLOT16}. These techniques have since evolved, with methods like ACoL that integrate CAM into an adversarial learning framework for object detection~\cite{DBLP:conf/cvpr/ZhangWF0H18}, and Puzzle-CAM which refines the quality of CAM-generated segmentation using reconstructive regularization loss~\cite{DBLP:conf/icip/JoY21}. The notion of pre-training segmentation models with explanations was further advanced by Pseudo Semantic Segmentation Labels (PSSL) which repurposes the explanations of classifier to annotate images for semantic segmentation pre-training~\cite{li2023distilling}.

As we strive to harness weakly-supervised pre-training for the specialized needs of medical image analysis, particularly segmentation, these varied approaches provide a rich tapestry of strategies to draw upon. They highlight the untapped potential of weak supervision not only as a tool for reducing reliance on extensive annotated datasets but also as a means to improve model interpretability and performance in a cost-effective manner. Our work builds on these insights, aiming to close the gap in the current landscape of weakly supervised methods for medical image segmentation and propel them to the forefront of pre-training methodologies.  

\section{Methods}
In this section, we introduce how the diagnosis-oriented localization labels are built to enable an end-to-end pre-training process for segmentation models. For each CXR image, DoLL localizes 14 sets of regions of interest via different pathological observations by distilling and boosting the explanations of weak classifiers. Applying this method to the massive images from CheXpert dataset, we pretrain the entire model and only finetune the segmentation adapters for various downstream settings with a unified backbone.

\subsection{DoLL: Diagnosis-orinted Localization Labels}\label{sec:method_gen}
As illustrated in Fig.~\ref{pipeline}, we present how the explanations of weak learners are used to construct the 14 sets of diagnosis-oriented labels. CheXpert~\cite{irvin2019chexpert} is a large-scale multi-label classification dataset for chest X-rays. It consists of 224,316 images of 65,240 patients labeled for the presence of 14 common chest radiographic observations such as ``Enlarged Cardiomediastinum'', ``Lung Opacity'', ``Pneumothorax'', ``Fracture'', ``Support Devices'', etc. Here, we select in total 191,027 frontal-view CXRs and train several multi-label classifiers for predicting the probability of each pathological observation.

Integrated gradient~\cite{DBLP:conf/icml/SundararajanTY17} is a gradient-based axiomatic attribution method for deep neural network, which requires almost no modification to the original network. By following a designed path from baseline to the input image, it computes in each step the integral of gradients, which reduces the unexpected noise on irrelevant pixels. Here we set the baseline as a black image with a linear path and apply the Riemann approximation for calculating the integral. $T$ denotes the total discretizing steps, $E_i^c$ the explanation result of integrated gradient for model $i$ while predicting label $c$, $\mathcal{L}^c_i$ the loss function of model $i$ for label $c$, and $X$ the input image. For each model $i$, we have:
\begin{align}
    E^c_i &= \frac{1}{T} \sum_{t=1}^{T} \frac{\partial \mathcal{L}^c_i(\frac{t}{T}X)}{\partial X}
     \overset{\rm T \to \infty}{\to} 
    \int _{\alpha=0} ^1 \frac{\partial \mathcal{L}^c_i(\alpha X)}{\partial X }d\alpha \quad .
\end{align}

\subsubsection{Boosted Distillation of Explanations}
The feature attribution map of a single model can be biased~\cite{DBLP:journals/corr/abs-2109-00707}. Here, the case is even more complicated, since we not only have multiple models but each of them predicts 14 observations with varying performances. In order to maximize the plausibility and faithfulness of our generated DoLL masks, we adopt a boosting strategy for distilling the helpful knowledge of each model when predicting each observation.

Assuming $M$ weak classifiers with $C$ observations, we first compute the weights $W_{m, c} \in \mathbb{R}^{M\times C}$ with $N$ samples according to their predicting performance on CheXpert, as illustrated in Algorithm~\ref{alg:boosting}.
\begin{algorithm}
\caption{Boosting the explanations for weak learners }\label{alg:boosting}
\begin{algorithmic}
\Require $y_n^c$ the CheXpert label of sample $n$ for observation $c$
\For{$c \in [1, C]$} \Comment{$C$ observations}
\State Initialize weights $S: S_1, S_2, \dots, S_N = \frac{1}{N}$
    \For{$m \in [1, M]$} \Comment{$M$ weak learners}
    \State Fit classifier $m$ to $N$ samples and obtain the predicted label $\hat{y}_n^c$ 
    \State $e_i = \frac{\sum_{n=1}^{N}S_n I(\hat{y}_n^c \neq y_n^c) }{\sum_{n=1}^{N}S_n}$ \Comment{Calculating the error}
    \State $W_{m, c} = \log(\frac{1-e_i}{e_i}) + \log(K-1)$ \Comment{weight for model $m$ on observation $c$}
    \State $S_n = S_n e^{W_{m, c} I(\hat{y}_n^c \neq y_n^c)}$ for $S_n \in S $
    \State $S_n = \frac{S_n}{\sum_n S_n}$ for $S_n \in S $ \Comment{Re-normalize $S$}
\EndFor
\EndFor
\State \Return $W$
\end{algorithmic}
\end{algorithm}
Moreover, we only consider the region of interest for the models with relatively high probability scores (beyond the threshold $\tau$) and have: 
\begin{align}
    N^c &= \{ \ i \ |  \ p_i^c > \tau, i \in \{ 1, \dots, M \} \ \}  \quad , \label{tau} \\ 
    \Tilde{E}^c &= \frac{1}{|N^c|} \sum_{m \in N^c}  W_{m, c} E^c_n \quad ,
\end{align}
where $p_i^c$ denotes the probability score of model $i$ for label $c$, $N^c$ the set of models with the probability score beyond the threshold $\tau$, and $|N^c|$ the cardinality of the set $N^c$.

At last, for each observation $c$, we binarize $\Tilde{E}^c$ with a threshold and convert the explanation result into Boolean values.
In this way, we obtain 14 segmentation labels for each chest radiograph which localize the discriminative pixels from different pathological views.

\begin{figure}[htbp]
    \begin{center}
        \includegraphics[width=\textwidth]{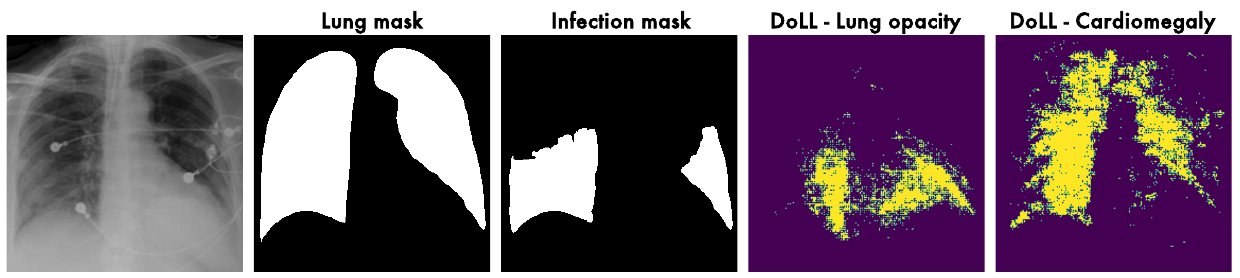}
    \end{center}
\caption{Example of our DoLLs and its ground truth masks}
\label{doll_dataset}
\end{figure}
Based on this method, we equip each frontal CXR in CheXpert with 14 sets of binary masks, where each describes the region of interest under a specific pathological perspective. Thereby, we obtain \textbf{a new large-scale dataset CheXpert-DoLL}, which enables an end-to-end pre-training for segmentation models. As presented in 
DoLLs under certain observations are coherent with ground truth masks for different segmentation tasks, indicating the advantage of pre-training with DoLLs in supporting a large variety of downstream settings.

\subsection{Downstream Segmentation Adapters}\label{sec:method_finetune}
For the pre-training stage, we pretrain both backbone and segmentation modules with DoLLs served as the segmentation labels. We minimize the binary cross entropy between DoLL and the output mask, that is
\begin{equation}
    \sum_{c=1}^{14} \ -d_{c, x} \ {\rm log}(p_{c, x}) - (1 - d_{c, x}) \ {\rm log}(1-p_{c, x})
\end{equation}
where $p_{c, x}$ is the probability and $d_{c, x}$ the DoLL for label $c$ on pixel $x$. 
The benefits of pre-training in an end-to-end manner is that the backbone and the segmentation module can be better correlated in their pretrained weights, thus decreasing the training difficulty in fine-tuning process. Besides, the strong clinical relevance and the specificity for CXR can help the model extract more general features and reach better performances on various downstream tasks.

Given both backbone and segmentation module pretrained, we apply them for various downstream tasks. As shown in Fig.~\ref{training}, we freeze the backbone and only update the segmentation module during fine-tuning. We name the finetuned modules as downstream segmentation adapters. The reasons for freezing the backbone are two-fold: First, only updating the parameters in the segmentation module demands less time and space for better training efficiency. Second, it adds more flexibility to implementations and requires less storage space. The model can be adapted to different segmentation tasks by only switching the adapter layers.

\section{Results}
In this section, we present the results of our work. We first introduce the experiment settings, and then present the overall comparison results between DoLL and baselines. Ablation and case studies have been done to confirm the advantages of DoLL. 

\subsection{Experiment Settings}
Here, we introduce DoLL generation from classification datasets, datasets used for pre-training and fine-tuning, baseline algorithms for comparisons.

\subsection{CheXpert-DoLL Generation}
 For the generation of CheXpert-DoLL, we train the following 17 models for a multi-label classification task on CheXpert: DenseNet121, MobileNetV2, AlexNet, ResNet50, ResNet50-32$\times$4d, ResNet101-32$\times$8d, Wide-ResNet50, Wide-ResNet101, Vgg16, ResNet152, ShuffleNetv2, DenseNet161, ShuffleNetv2-x2, ResNet101, EfficientNetB4, MobileNetV3 and ViT-base-32, with the mean AUC around 0.8 for each model. As shown in Fig.~\ref{pipeline}, the overall idea of DoLL generation is to first aggregate the explanation results (e.g., sailency maps obtained by Integrated Gradients) extracted from these well-trained classifiers, and then to establish the ``consensus'' of visual explanations among these classifiers as the pseudo labels for semantic segmentation tasks~\cite{li2023distilling,li2023cross}. Please refer to Section~\ref{sec:method_gen} for details on the algorithms for generating DoLL on images. In general, the greater number of models and the better classification performance can definitely improve its exactness for the DoLL generation, and lead to better pre-training results. For a compromise between time consumption and exactness, we suggest the number of models to be set around 15-20. 
 
 Then, we generate the CheXpert-DoLL over these observations with the total number of models $M=17$, the threshold $\tau = 0.1$, the $80$-th percentile for the binarizing threshold, the steps in IG $T=5$, and $K=3$ for the boosting. A larger value of binarizing threshold leads to a more concentrated region of interest, and vice versa for a smaller value. Based on our empirical findings, more concentrated DoLL can improve the segmentation performance for the lung-related regions but degrade the model's generalization ability, leading to worse performance for the segmentation of heart, clavicles, supporting devices, etc. 

\begin{table}[htbp]
\centering
\caption{Overview of datasets used in our experiments}
\begin{tabular}{@{}c|c|ccc|cc@{}}
\toprule
\multicolumn{1}{c|}{Datasets} & Task                            & Train   & Val  & Test & Purpose \\ \midrule\midrule
CheXpert (frontal)             & Multi-label classification      & 191,027 & -    & -    &     pre-training            \\ \midrule
\multirow{2}{*}{COVID-QU-EX}   & Lung segmentation               & 7658    & 1903 & 2395 &     fine-tuning      \\
                               & COVID-19 infection segmentation & 1864    & 466  & 583  &        fine-tuning       \\ \midrule
JSRT                           & Multi-organ segmentation        & 171     & 25   & 50   &           fine-tuning        \\ \bottomrule
\end{tabular}
\label{dataset}
\end{table}

\subsubsection{Segmentation Tasks}
In the experiments, we evaluate the pre-training methods on three downstream tasks: lung segmentation on COVID-QU-EX~\cite{DBLP:journals/cbm/TahirCKRQKKIRAM21} (Lung), COVID-19 infectious region segmentation on COVID-QU-EX (Infection) and multi-organ segmentation including lungs, heart and clavicles on Chest X-ray Landmark Segmentation Dataset~\cite{Gaggion_2022} (JSRT). 
\textbf{COVID-QU-EX} is a dataset collected by the researchers of Qatar University. It is composed of X-ray images of the human chest labelled as either
``Healthy'', ``COVID-19'', or ``Pneumonia''. It provides as well the corresponding ground-truth infection segmentation masks and lung segmentation masks. In our experiments, we use the lung masks for lung segmentation tasks, and the infection masks for a direct segmentation from X-ray images without mediating any lung mask.
\textbf{Chest x-ray Landmark Segmentation Dataset} contains 911 landmark annotations for chest X-ray images from JSRT, Shenzhen, Montgomery and Padchest datasets. Here, we use the subset with \textbf{JSRT} images for the multi-organ segmentation of ``Right Lung'', ``Left Lung'', ``Heart'', ''Left Clavicle'', and ``Right Clavicle''. More detailed descriptions of these datasets can be found in Table~\ref{dataset}. Also, please refer to Section~\ref{sec:method_finetune} for details on the algorithms for fine-tuning CheXpert-DoLL pre-trained models for segmentation tasks. 

\subsubsection{Baselines and Setups}
To validate our DoLL pre-training strategies, we propose the following pre-training baselines. We consider both supervised pre-training with large-scale classification datasets for natural and medical images, and self-supervised pre-training strategies.
\begin{itemize}
    \item \textbf{ImageNet}: we finetune the entire 3-channel model on target datasets with initialization from officially-released pretrained backbone weights;
    
    \item \textbf{CheXpert}~\cite{irvin2019chexpert}: the 3-channel backbone is pretrained with a multi-label classification task on CheXpert and we finetune the entire model for downstream tasks;
    
    \item \textbf{MoCo-CXR}~\cite{DBLP:conf/midl/SowrirajanYNR21}: it is an adaptation of Momentum Contrast (MoCo) to produce better representation and initialization for the detection of pathologies in chest X-rays. We initialize the backbone with the officially-released pretrained weights in MoCo-CXR, and finetune the entire model for the downstream tasks;

    \item  \textbf{MoCo-v2}~\cite{DBLP:journals/corr/abs-2003-04297}: the method implements the SimCLR's design in the MoCo framework, by using an MLP head and better data augmentation strategies. We here initialize the backbone with the officially-released pretrained weights on ImageNet, and finetune the entire model for the downstream tasks;

    \item \textbf{GrayScale ImageNet}~\cite{DBLP:conf/eccv/XieR18}: we adopt both the official released single-channel backbone structure and weights, and finetune the entire model;

    \item \textbf{Scratch}: for segmentation models without pretrained backbones such as U-Net, we train the model directly on the target dataset. 
\end{itemize}

For a more fair comparison, we propose \textbf{DoLL} with 3-channel model input, and \textbf{DoLL-1channel} for single channel. 
We select three representative segmentation network architectures: the classic PSPNet~\cite{DBLP:conf/cvpr/ZhaoSQWJ17}-ResNet50~\cite{DBLP:conf/cvpr/HeZRS16}, the Transformer-based Segformer-Mix Transformer~\cite{DBLP:conf/nips/XieWYAAL21}, and the U-Net~\cite{DBLP:conf/miccai/RonnebergerFB15} which is designed for medical image segmentation. The segmentation performance is evaluated via the metrics including the \textit{mean Intersection over Union} (mIoU), \textit{mean Accuracy} (mAcc) and \textit{Dice Similarity Coefficient} (Dice). For the pre-training, we train 30 epochs for each model with learning rate 0.01, batch size 128, input image size 224 and data augmentation methods including random crops, flipping and rotations. For the fine-tuning, we train each model or adapter for 20,000 iterations, and select the checkpoint with highest IoU on validation set.

\subsection{Overall Results}
We present in Table~\ref{overall} the results of baselines and ours on all the downstream tasks. By only updating the segmentation module, our method can reach better performance than all the other entirely finetuned baselines which are entirely finetuned, including the large-scale classification datasets and self-supervised learning strategies. Moreover, our methods show great improvements regardless of different model architectures. For COVID-19 infection segmentation, the outperformance
is more evident, which indicates that pre-training with DoLL reaches a deeper extraction of visual features for CXRs, not only just in the contour, but also the pathological interpretations. 

\begin{figure}
    \centering
    \includegraphics[width=0.8\textwidth]{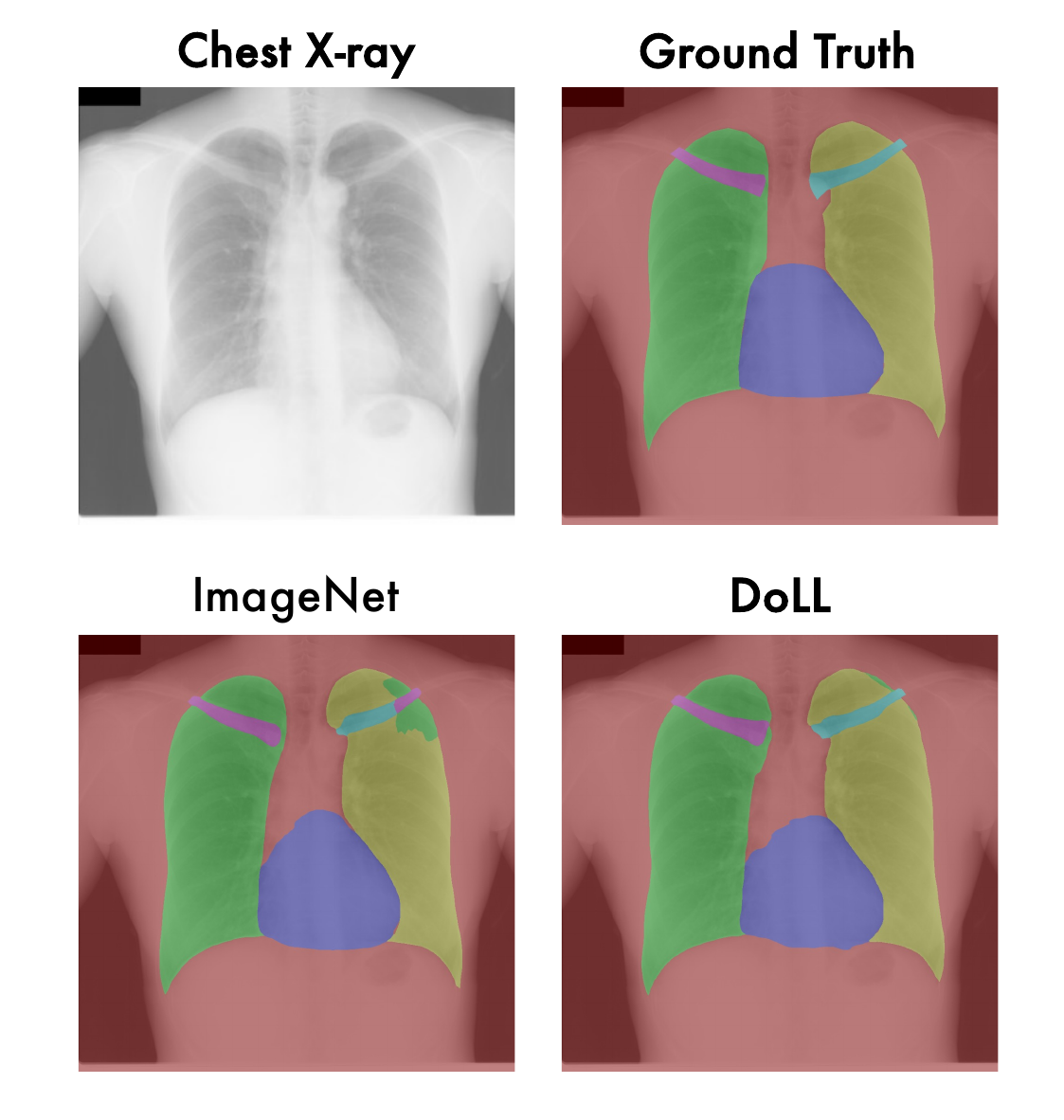}
    \caption{Visualization of segmentation results on JSRT for ImageNet and DoLL pretrained models}
    \label{seg}
\end{figure}

We have also conducted experiments with fixed backbones for the baseline methods. Most of the metrics get worse compared to the entire fine-tuning. This demonstrates that our pre-trained weights are better correlated between backbone and segmentation module.
As for the visualization of segmentation, we provide in Fig.~\ref{seg} an example of multi-organ segmentation from JSRT dataset. We observe that our method is closer to the ground-truth segmentation mask, and shows significant improvement in segmenting clavicles. 


\begin{table}[htbp]
\centering
\caption{Overall results on testing set for COVID-19 infection segmentation (COVID-19), lung segmentation (Lung) and multi-organ segmentation (JSRT)}
\begin{tabular}{@{}ccccccccc@{}}
\toprule
\multirow{3}{*}{Model} &
  \multirow{3}{*}{pre-training} &
  \multicolumn{2}{c}{\multirow{2}{*}{COVID-19}} &
  \multicolumn{2}{c}{\multirow{2}{*}{Lung}} &
  \multicolumn{3}{c}{\multirow{2}{*}{JSRT}} \\
 &
   &
  \multicolumn{2}{c}{} &
  \multicolumn{2}{c}{} &
  \multicolumn{3}{c}{} \\ \cmidrule(l){3-9} 
 &
   &
  IoU &
  \multicolumn{1}{c|}{Acc} &
  IoU &
  \multicolumn{1}{c|}{Acc} &
  mIoU &
  mAcc &
  mDice \\ \midrule\midrule
\multicolumn{1}{c|}{\multirow{7}{*}{\begin{tabular}[c]{@{}c@{}}PSPNET\\ -ResNet50\end{tabular}}} &
  \multicolumn{1}{c|}{ImageNet} &
  66.74 &
  \multicolumn{1}{c|}{80.05} &
  89.37 &
  \multicolumn{1}{c|}{94.39} &
  86.54 &
  90.84 &
  92.65 \\
\multicolumn{1}{c|}{} &
  \multicolumn{1}{c|}{CheXpert} &
  72.73 &
  \multicolumn{1}{c|}{84.21} &
  90.7 &
  \multicolumn{1}{c|}{95.12} &
  85.56 &
  89.69 &
  92.00 \\
\multicolumn{1}{c|}{} &
  \multicolumn{1}{c|}{MoCo-CXR} &
  72 &
  \multicolumn{1}{c|}{83.72} &
  94.83 &
  \multicolumn{1}{c|}{97.34} &
  89.96 &
  92.70 &
  94.62 \\
\multicolumn{1}{c|}{} &
  \multicolumn{1}{c|}{MoCo-v2} &
  72.39 &
  \multicolumn{1}{c|}{83.98} &
  94.96 &
  \multicolumn{1}{c|}{97.41} &
  89.78 &
  92.85 &
  94.49 \\
\multicolumn{1}{c|}{} &
  \multicolumn{1}{c|}{GrayScale ImageNet} &
  69.75 &
  \multicolumn{1}{c|}{82.18} &
  91.67 &
  \multicolumn{1}{c|}{95.65} &
  89.13 &
  93.18 &
  94.09 \\ \cmidrule(l){2-9} 
\multicolumn{1}{c|}{} &
  \multicolumn{1}{c|}{\bf DoLL} &
  77.48 &
  \multicolumn{1}{c|}{87.31} &
  \textbf{95.31} &
  \multicolumn{1}{c|}{\textbf{97.60}} &
  \textbf{90.98} &
  \textbf{93.96} &
  \textbf{95.20} \\
\multicolumn{1}{c|}{} &
  \multicolumn{1}{c|}{\bf DoLL-1channel} &
  \textbf{79.3} &
  \multicolumn{1}{c|}{\textbf{88.45}} &
  94.53 &
  \multicolumn{1}{c|}{97.19} &
  90.03 &
  93.31 &
  94.63 \\ \midrule\midrule
\multicolumn{1}{c|}{\multirow{3}{*}{Segformer-MiT}} &
  \multicolumn{1}{c|}{ImageNet} &
  72.13 &
  \multicolumn{1}{c|}{83.81} &
  93.97 &
  \multicolumn{1}{c|}{96.89} &
  \multicolumn{1}{l}{93.28} &
  \multicolumn{1}{l}{95.14} &
  \multicolumn{1}{l}{96.49} \\
\multicolumn{1}{c|}{} &
  \multicolumn{1}{c|}{CheXpert} &
  83.67 &
  \multicolumn{1}{c|}{91.11} &
  94.46 &
  \multicolumn{1}{c|}{97.15} &
  \multicolumn{1}{l}{93.09} &
  \multicolumn{1}{l}{95.32} &
  \multicolumn{1}{l}{96.38} \\ \cmidrule(l){2-9} 
\multicolumn{1}{c|}{} &
  \multicolumn{1}{c|}{\bf DoLL} &
  \textbf{92.12} &
  \multicolumn{1}{c|}{\textbf{95.9}} &
  \textbf{95.91} &
  \multicolumn{1}{c|}{\textbf{97.91}} &
  \multicolumn{1}{l}{\textbf{93.66}} &
  \multicolumn{1}{l}{\textbf{95.72}} &
  \multicolumn{1}{l}{\textbf{96.69}} \\ \midrule\midrule
\multicolumn{1}{c|}{\multirow{2}{*}{U-Net$^*$}} &
  \multicolumn{1}{c|}{Scratch} &
  75.42 &
  \multicolumn{1}{c|}{85.99} &
  94.15 &
  \multicolumn{1}{c|}{96.99} &
  70.64 &
  80.44 &
  82.32 \\ \cmidrule(l){2-9} 
\multicolumn{1}{c|}{} &
  \multicolumn{1}{c|}{\bf DoLL} &
  \textbf{81.05} &
  \multicolumn{1}{c|}{\textbf{89.54}} &
  \textbf{96.54} &
  \multicolumn{1}{c|}{\textbf{97.75}} &
  \textbf{74.24} &
  \textbf{82.84} &
  \textbf{84.87} \\ \bottomrule
\end{tabular}
\label{overall}
\end{table}

\subsubsection{Adaptation for Downstream Tasks}
In our experiments, we consider a great variety of downstream settings to validate our pre-training method. Besides the common lung segmentation, we include as well infectious region segmentation and few-shot multi-organ segmentation. 
The results well support the conclusion that by generating labels via the 14 observations, our pretrained model reaches a better generalization ability and a deeper understanding to the CXR images, leading to an all-round enhancement on various downstream segmentation tasks.

\begin{figure}[htbp]
    \centering
    \includegraphics[width=\textwidth]{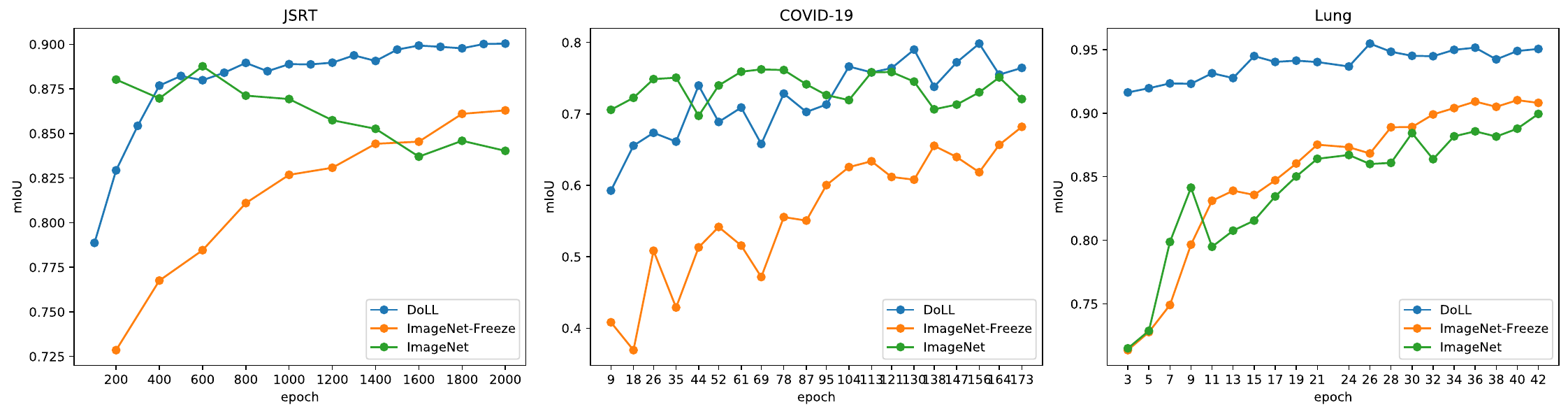}
    \caption{Record of mIoU on validation set during the PSPNet fine-tuning process for JSRT, COVID-19 and Lung segmentation tasks}
    \label{fig}
\end{figure}

\subsubsection{Efficiency}
The freezing of backbone during the fine-tuning process, i.e. the downstream segmentation adapters, not only facilitates the future implementations by requiring less storage space and less time for loading and switching, but also improves the training efficiency. As illustrated in Fig.~\ref{fig}, we record the mIoU score on validation set for each epoch. For all the downstream settings, our pretrained model presents a stable performance and fast convergence. By contrast, ImageNet pre-training shows an easy tendency of overfitting under few-shot setting (JSRT) and slow convergence with lung segmentation. Moreover, we freeze the backbone for ImageNet pre-training as well, whereas the results are not as satisfactory as ours. First, the less powerful backbone limits its performance on downstream tasks. Besides, the random initialization of segmentation module leads to a broken relation between backbone and segmentation module. Failing to get correlated adds extra difficulty to the fine-tuning process, since the model not only needs to adapt into downstream tasks, but also to build correlations between backbone and segmentation module. 

\subsection{Ablation Study}
We conduct an ablation study to validate the effectiveness of the boosting module in our proposed annotating method. Table~\ref{ablation} presents the IoU results on our three downstream segmentation settings with PSPNET-ResNet50. We average the results from different explanation methods instead of adopting the boosting coefficients. The results demonstrate the advantages of boosting weak learners with improvements in all metrics.

\begin{table}[htbp]
\centering
\caption{IoU results of ablation study with PSPNET-R50}
\begin{tabular}{@{}c|ccccccc@{}}
\toprule
\multirow{2}{*}{\quad PSPNET-R50 \quad } & \multirow{2}{*}{\quad COVID-19} & \multirow{2}{*}{\quad Lung\quad } & \multicolumn{5}{c}{JSRT}                                                           \\ \cmidrule(l){4-8} 
         &       &       & RL    & LL    & H     & RCLA  & LCLA  \\ \midrule\midrule
Averaged & 76.10 & 95.01 & 95.73 & 94.99 & 93.00 & 85.51 & 84.36 \\
Boosted                     & \textbf{77.48}            & \textbf{95.31}        & \textbf{96.07} & \textbf{95.59} & \textbf{93.06} & \textbf{85.66} & \textbf{84.50} \\ \bottomrule
\end{tabular}
\label{ablation}
\end{table}

\subsection{Case Studies}
Here, we use case studies to analyze the time consumption of DoLL and the potential impact of explanation accuracy in DoLL to the overall performance.
\subsubsection{Time Cost}
Questions may come up if our method costs too much time compared to the conventional backbone pre-training pipeline, since it involves the generation of DoLL, the end-to-end pre-training, and the downstream fine-tuning process. First, the generation of DoLL does not cost much time and effort. By using 17 classification models with mean AUC around 0.8, we can already construct the DoLL with pretty high quality. The training of these classification models can be easily achieved without the necessity of bringing special tricks. Second, we would like to address that both the DoLL and pretrained checkpoints will be released publicly. We demonstrated in the previous section that pre-training with the DoLL can largely improve the efficiency during fine-tuning, costing less time and space. Besides, once the DoLLs are generated, it can be directly applied for other pre-training tasks. Even if the desired network architecture is not involved in the released checkpoints, the end-to-end pre-training will not cost much more time than the conventional pre-training pipeline, and largely improve the efficiency in the fine-tuning process.

\subsubsection{What if the explanations are inaccurate ?}
We have adopted several techniques to guarantee the faithfulness of generated DoLLs. As mentioned, we distill and boost the explanations of weak learners to avoid the biasedness from misclassification. Besides, increasing the number of classifiers, improving their classifying performances, tuning the threshold $\tau$ in Equation~\eqref{tau}, and modifying the binarizing threshold can all be helpful to further improve the exactness. In our experiment, we generate DoLLs with 17 models and pretrain the entire segmentation model for 30 epochs without any special designs. As a result, we have achieved very satisfactory performance on downstream segmentation tasks, with evident improvements compared to previous pretrainig pipelines.

\section{Discussions}
 This work introduces DoLL, an innovative pre-training method for medical image segmentation specifically tailored for chest X-rays. The extensive CheXpert-DoLL dataset, generated using interpretations of classifiers trained for 14 pathological observations with CheXpert, significantly boost model performance on tasks such as lung and COVID-19 infection segmentation, and few-shot multi-organ segmentation. Compared to traditional pre-training that only applies to backbone weights, DoLL enables end-to-end pre-training for both backbone and segmentation modules, effectively reducing the need for large annotated datasets. The resulting approach not only deepens feature extraction capabilities but also generalizes better to new tasks, allowing for the backbone to be fixed during fine-tuning processes. As part of contribution, we have compiled and released the CheXpert-DoLL dataset, set to catalyze future advancements in chest X-ray segmentation research and clinical applications.  



Despite the notable successes, there remain open research issues that invite further exploration.
\begin{itemize}
    \item \textbf{Generalizability.} One significant area is the generalizability of DoLL across other imaging modalities beyond chest radiographs. As medical imaging encompasses a wide array of modalities - each with its own unique set of characteristics and challenges - future research could investigate the applicability and adaptation of DoLLs to CT scans, MRI, and ultrasound images. Some earlier work that leverages 2D images to pre-train 3D models actually shed the light to this area~\cite{DBLP:conf/isbi/ZhangWBLXDX23}.
    
    \item \textbf{Robustness.} Another open issue pertains to the robustness of the generated DoLL against variations in pathology presentations and imaging conditions. While the paper details the success in uncovering 14 pathological observations, the diversity of pathological conditions and their manifestations in images suggest the need to validate and possibly enhance the robustness of the extracted explanations against such variations.
    %
    Furthermore, there is the potential to expand the utility of DoLL to enhance not just segmentation tasks but also broader aspects of computer-aided diagnosis such as prognostic modeling and personalized medicine, which could significantly impact patient care.
    
    \item \textbf{Interpretability.} In terms of the technical aspects, there lies an avenue for research in optimizing the extraction and utilization of classifier explanations. For instance, DoLL are derived using Integrated Gradients, but further research could explore alternative methods such as LIME, SHAP or their variants~\cite{li2023g}, potentially providing an enriched set of localization labels that could further refine the pre-training process.
    
    \item \textbf{Ethics.} Lastly, ethical considerations and biases within AI applications in medicine necessitate ongoing research. Ensuring that the algorithms are fair, transparent, and equitable across different populations is a multifaceted challenge that intersects with data diversity, algorithmic design, and regulatory compliance.
\end{itemize}
The work undertaken has laid a robust foundation that can catalyze future explorations within these open research areas. By addressing these challenges, the scientific community can further the advancements in medical image analysis, ultimately contributing to the enhancement of patient outcomes and the optimization of healthcare delivery.

%
%
%

\bibliography{mybibliography}
\clearpage

\appendix

\end{document}